\documentclass[a4paper,11pt]{article}

\usepackage{jheparxiv}

\usepackage{mathrsfs}
\usepackage{bbm}
\usepackage{bbold}
\usepackage{color}
\usepackage{slashed}
\usepackage{tikz}
\usepackage{parskip}
\usepackage[nice]{nicefrac}

\newcommand{\lambdabar}{{\mkern0.75mu\mathchar '26\mkern -9.75mu\lambda}}
\newcommand{\nf}[2]{\nicefrac{#1}{#2}}
\newcommand{\eqnref}[1]{Eq. (\ref{#1})}

\newcommand{\be}{\begin{equation}}
\newcommand{\ee}{\end{equation}}
\newcommand{\bi}{\begin{itemize}}
\newcommand{\ei}{\end{itemize}}
\newcommand{\bea}{\begin{eqnarray}}
\newcommand{\eea}{\end{eqnarray}}
\newcommand{\tr}{\text{tr}\,}
\newcommand{\ps}{\tsf{-}}
\newcommand{\sca}{\tsf{+}}

\newcommand{\nn}{\nonumber}

\newcommand{\vphi}{\varphi}
\newcommand{\vkap}{\varkappa}
\newcommand{\eps}{\varepsilon}

\newcommand{\Ai}{\trm{Ai}}
\newcommand{\F}{F}
\newcommand{\Fs}{F_{\tsf{\tiny{Q}}}}
\newcommand{\mphi}{m_{\phi}}
\newcommand{\defto}{:=}

\newcommand{\mbf}[1]{\mathbf{#1}}
\newcommand{\trm}[1]{\mbox{\textrm{#1}}}

\newcommand{\figref}[1]{Fig. \ref{#1}}
\newcommand{\figrefa}[1]{Fig. \ref{#1}a}
\newcommand{\figrefb}[1]{Fig. \ref{#1}b}

\newcommand{\e}{\mathbb{e}}
\newcommand{\K}{\tsf{K}}

\newcommand{\eqnrefs}[2]{Eqs. (\ref{#1}) and (\ref{#2})}

\newcommand{\tsf}[1]{\textsf{#1}}

\definecolor{stdout}{rgb}{0.6,0.2,0.2}
\definecolor{gravy}{rgb}{0.95,0.9,0.9}
\definecolor{dgreen}{rgb}{0.05,0.6,0.05}
\definecolor{dgreenb}{rgb}{0.05,0.8,0.05}

\newcommand{\Ben}[1]{\textcolor{red}{#1}}

\title{\boldmath Axion-like-particle decay in strong electromagnetic backgrounds}

\author[a,1]{B. King,\note{Corresponding author.}}
\author[b]{B. M. Dillon,}
\author[c]{K. A. Beyer,}
\author[c]{G. Gregori}

\affiliation[a]{Centre for Mathematical Sciences, University of Plymouth, Plymouth, PL4 8AA, United 
Kingdom}
\affiliation[b]{Jo\u{z}ef Stefan Institute, Jamova 39, 1000 Ljubljana, Slovenia}
\affiliation[c]{Department of Physics, University of Oxford, Parks Road, Oxford, OX1 3PU, United Kingdom}

\emailAdd{b.king@plymouth.ac.uk}

\abstract{The decay of a massive pseudoscalar, scalar and U(1) boson into an electron-positron pair in the presence of strong electromagnetic backgrounds is calculated. Of particular interest is the constant-crossed-field limit, relevant for experiments that aim to measure high-energy axion-like-particle conversion into electron-positron pairs in a magnetic field. The total probability depends on the quantum nonlinearity parameter - a product of field and lightfront momentum invariants. Depending on the seed particle mass, different decay regimes are identified. In the below-threshold case, we find the probability depends on a non-perturbative tunneling exponent depending on the quantum parameter and the particle mass. In the above-threshold case, we find that when the quantum parameter is varied linearly, the probability oscillates nonlinearly around the spontaneous decay probability. A strong-field limit is identified in which the threshold is found to disappear. In modelling the fall-off of a quasi-constant-crossed magnetic field, we calculate probabilities beyond the constant limit and investigate when the decay probability can be regarded as locally constant.}

\begin{document}

\maketitle
\flushbottom

\section{Introduction}\label{sec:intro}
``Naturalness'' seems incompatible with the Standard Model (SM) when one considers the ``strong-CP'' problem, which asks why charge-parity (CP) conjugation invariance is violated so little in the strong sector despite an explicit CP-violating term in the QCD Lagrangian (this would induce a large but unobserved neutron electric dipole moment). An attractive solution is the Peccei-Quinn (PQ) mechanism which promotes the CP-violating term to be a dynamical parameter that can relax to zero and predicts the existence of a pseudoscalar Nambu-Goldstone boson called the axion \cite{Peccei:1977hh}, which has a weak coupling to photons and electrons, as well as other SM particles.  Other beyond-the-Standard-Model scenarios predict the existence of light bosonic states that couple weakly to photons and electrons, which are referred to collectively as Axion-Like-Particles (ALPs). 
They have subsequently been suggested to explain various astrophysical phenomena such as the transparency of the universe to high energy gamma-rays \cite{angelis07,mirizzi08,simet08,conde09}, and the 3.55 keV galaxy cluster emission line \cite{bulbul14,boyarsky14,jaeckel14}.

A promising route to detecting ALPs is through their coupling to SM particles in the electromagnetic sector. The coupling of ALPs to the electromagnetic field is exploited in Light-Shining-through-the-Wall (LSW) experiments (for a review see \cite{Redondo:2010dp}) to convert laser photons in a magnetic background into ALPs, which then propagate through a ``wall'' and into a low-noise detection region with a background magnetic field. ALPs in the presence of this background field are then reconverted into photons, which provide the experimental signal. The current state-of-the-art LSW experiment is the ALPS I experiment \cite{Ehret:2010mh}, however upgrades to this set-up and other more advanced LSW experiments are planned for the future \cite{Bahre:2013ywa,Capparelli:2015mxa}.
Helioscope experiments, i.e. CAST \cite{cast17} and the proposed IAXO experiment \cite{Armengaud:2014gea}, also use a similar detection set-up, but since the generation stage occurs in complex astrophysical environments such as in the sun, both production via the di-photon coupling in e.g. the Primakoff process, and production via the electron-ALP coupling in e.g. axionic-Compton emission \cite{borisov96}, is being measured.
This means that the signal in helioscope experiments, unlike LSW, is also sensitive to the coupling of ALPs to electrons, and a bound on this coupling has been derived by the CAST collaboration \cite{Barth:2013sma}.

In the current paper we study the process of an ALP decaying to an electron-positron pair in a high-intensity electromagnetic field via a direct coupling of the ALP to electrons. This adds to the overall discussion on altering particle decay properties using external electromagnetic (EM) fields and could be of interest for future lab-based ALP searches. In addition to the decay of photons in magnetic \cite{Toll:1952rq, klepikov54, baier07} and plane-wave \cite{heinzl10,titov12,nousch12} fields, the decay lifetime of a muon has also been investigated, and shown to be only slightly changed in an EM background in \cite{ritus69,becker83}, (more recent speculations to the contrary were criticised in the literature \cite{narozhny08}). Depending on the set-up of background fields, a magnetic field may enhance or suppress particle production. For example, for constant homogeneous parallel electric and magnetic fields in QED \cite{bunkin70} and scalar QED \cite{popov72}, a weak magnetic field has been found to slightly enhance the decay of the vacuum into electron-positron pairs (more details can be found in the review \cite{dunne04b}), but for parallel Sauter type electric and magnetic fields \cite{Sogut:2017ksu} and for the decay of a neutral scalar to two charged scalars in a thermal bath \cite{piccinelli17}, to suppress particle production.

The production of ALPs via their coupling to photons in a circularly-polarised laser beam has been studied in \cite{villalbachavez12}.
The production of ALPs in the interactions between electrons and high-intensity electromagnetic fields has been studied previously in \cite{burton18, king18b,king18c,king18e}, with ALP-seeded electron-positron pair production in a monochromatic laser background also being considered in \cite{king18b}.
These papers demonstrated how lab-based experiments using high-intensity lasers and electrons may provide lab-based bounds on the ALP electron coupling, complementary to those derived from helioscope experiments.
The current paper extends this work by considering the decay of ALPs to electron-positron pairs in quasi-constant magnetic fields, derived as a limit of the case in which the process occurs in a plane-wave electromagnetic background.

This process is relevant for both terrestrial experiments utilising strong magnetic fields for ALP conversion and searches for extraterrestrial ALPs from strongly-magnetised objects \cite{harding06}. To perform calculations in strong electromagnetic backgrounds, we employ the Furry picture \cite{furry51}. Solutions to the Dirac equation in a plane-wave electromagnetic background, so-called ``Volkov'' states, represent the fermions ``dressed'' in the external electromagnetic field \cite{volkov35}. As such, the derivation of ALP decay rates in quasi-constant electromagnetic fields has much in common with high-intensity QED (reviews can be found in \cite{ritus85,marklund06,gies09,dipiazza12b,narozhny15}), with the decay of photons in a laser background being measured experimentally in the E144 experiment \cite{burke97,bamber99}. Due to the immense number of laser photon ``probe'' particles, most interest in extensions of high-intensity QED has been in the ALP-diphoton coupling. 
This can manifest itself in the polarisation properties of a photon probe \cite{doebrich10,villalbachavez14,villalbachavez17} or through parametric excitation \cite{mendonca07}. Recent calculations have begun exploring the possibility of using the collision of electron bunches with laser pulses to measure the ALP-electron coupling, for example using weak \cite{king18c}, strong \cite{king18c} and intermediate many-cycle \cite{king18b} laser pulses, or leveraging collective effects such as coherent emission \cite{king18e}.

The paper is structured as follows. In Sec. \ref{planewave}, we present an example derivation of ALP decay in a plane-wave electromagnetic background with finite support (e.g. a laser pulse), focusing on the decay of a massive pseudoscalar to an electron-positron pair. Derivations for a scalar and a vector boson follow a very similar format and final results for these cases are presented. We then perform a local expansion of the probability and obtain what is often referred to as the ``locally-constant-field-approximation'' (LCFA) \cite{king15d,dipiazza18a,king18d}. In Sec. \ref{constantfields} we analyse the constant crossed field (CCF) result, which is integrated over the non-constant background to form the LCFA. Asymptotic and perturbative limits for below- and above-threshold decay are presented, as is a description of how the non-perturbative pair-creation tunneling and oscillation exponents depend upon the ALP mass, particle energy and field strength. In Sec. \ref{edgeeffects} we consider the effect of the detector's magnetic field beyond the LCFA and show how large field gradients at the detector edge change the interpretation of a local production of pairs and can influence the total yield. In Sec. \ref{discussion} we discuss the results and conclude.

\section{Derivation of pseudoscalar decay probability in a plane-wave pulse} \label{planewave}
We begin by considering a pseudoscalar particle, $\phi$, with four-momentum $k$ and mass $m_{\phi}$, decaying to an electron-positron pair in a plane-wave electromagnetic background. The scaled vector potential $a^{\mu} = eA^{\mu}(\vphi)$, where $e$ is the charge of a positron and $A^{\mu}$ the vector potential, depends on a single variable, the phase $\vphi=\vkap\cdot x$ of the background. The lightlike wavevector of the background $\vkap$ is transverse to the vector potential, $\vkap\cdot a = 0$, and we represent $a^{\mu}(\vphi) = m\xi(\vphi) \eps^{\mu}$, where $\xi(\vphi)=\xi g(\vphi)$ is the local classical intensity parameter, $g(\vphi)$ is the pulse envelope ($|g(\vphi)|\leq 1$), $\eps^{\mu} = (0,\pmb{\eps}^{\perp},0)^{\mu}$ is the polarisation and $m$ is the electron mass. We choose a system of co-ordinates in which $\vkap = (\vkap^{+}/2)(1,0,0,1)$ so that $\vphi = \vkap^{+}x^{-}/2=\vkap^{0}x^{-}$, and use lightfront co-ordinates: $x^{\pm} = x^{0} \pm x^{3}$, $x^{\perp} = (0,x^{1},x^{2},0)$, $x_{\pm} = x^{\mp}/2$, $x_{\perp} = -x^{\perp}$. We will define the seed particle's mass parameter $\delta$, through $\delta^{2} = k^{2}/m^{2}$. 
\newline

\begin{figure}[h!!]
\centering
\includegraphics[width=3cm]{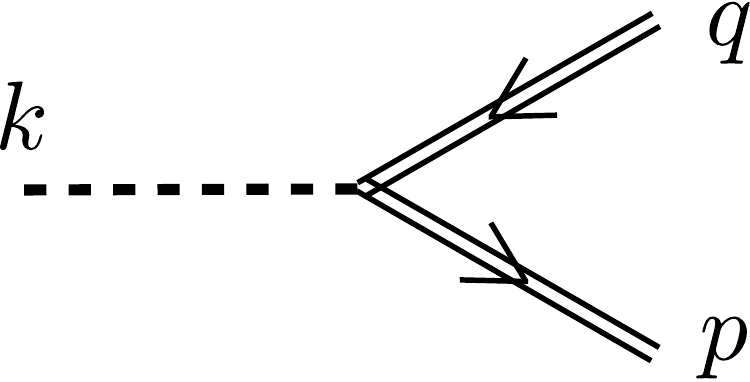}
\caption{Feynman diagram for pseudoscalar-seeded electron-positron pair production.}\label{fig:fd}
\end{figure}

To calculate the probability of decay in the electromagnetic background, we assume the produced fermions are solutions to the Dirac equation in a plane-wave electromagnetic background (Volkov states). This is depicted by the double fermion line in \figref{fig:fd}.
The scattering-matrix element for a pseudoscalar is then
\bea
\tsf{S}_{\tsf{fi}} = ig_{\phi e}~\int d^{4}x~\phi~ \overline{\psi}_{p}\gamma_{5} \psi_{q}^{+}, \label{eqn:Sfi1}
\eea
where $g_{\phi e}$ is the electron-pseudoscalar coupling, with Volkov states:
\bea
\overline{\psi}_{p} = \overline{E}_{p}(\vphi)\,\frac{\overline{u}_{r}(p)}{\sqrt{2p^{0}V}}\e^{ip\cdot x + iS_{p}(\vphi)},\qquad
\psi_{q}^{+} = E_{-q}(\vphi)\,\frac{v_{r'}(q)}{\sqrt{2q^{0}V}} \e^{iq\cdot x + iS_{-q}(\vphi)},
\eea
\bea
E_{p}(\vphi) = 1_{4}+ \frac{\slashed{\vkap}\slashed{a}(\vphi)}{2\,\vkap\cdot p}, \qquad S_{p}(\vphi) = \int^{\vphi} \frac{2 p\cdot a(\psi) - a^{2}(\psi)}{2\,\vkap\cdot p}~d\psi, \nn
\eea
where we assume that the pseudoscalar is in a plane-wave state, $\phi = \e^{-ik\cdot x}/\sqrt{2k^{0}V}$, has mass $m_{\phi}$, and $V$ is a normalisation volume. 

Integrating \eqnref{eqn:Sfi1} over the ``$-$'' and ``$\perp$'' co-ordinates we have
\bea
\tsf{S}_{\tsf{fi}} = \frac{i g_{\phi e} (2\pi)^{3}}{\vkap^{0}\sqrt{8p^{0}q^{0}k^{0}V^{3}}}\delta^{\perp,-}\left(p+q-k\right) \int d\vphi~\e^{i\vphi r + iS_{p}(\vphi) + iS_{-q}(\vphi)}~\overline{u}_{r}(p)\overline{E}_{p}(\vphi) \gamma^{5} E_{q}(\vphi)v_{r'}(q),\nn \\
\eea
where a measure of the lightfront momentum absorbed from the electromagnetic background is given by 
\[
r= \frac{p^{+} + q^{+} - k^{+}}{2\vkap^{0}}.
\]
To obtain the probability we must square the matrix element
\bea
\sum_{\tsf{spin}}\tr|\tsf{S}_{\tsf{fi}}|^{2} &=& \left[\frac{g_{\phi e} (2\pi)^{3}\,\delta^{\perp,-}\left(p+q-k\right)}{\vkap^{0}\sqrt{8p^{0}q^{0}k^{0}V^{3}}}\right]^{2} 
\int d\vphi\,d\vphi'\,\tsf{T}\,\e^{i r(\vphi-\vphi') + i\left[S_{p}(\vphi)-S_{p}(\vphi')\right] + i\left[S_{-q}(\vphi)-S_{-q}(\vphi')\right]}, \nn \\ \label{eqn:sfi6}
\eea
where the trace terms are included in the factor
\[
\tsf{T} = \sum_{r,r',s,s'}\tr \overline{u}_{r}(p)\overline{E}_{p}(\vphi) \gamma^{5} E_{q}(\vphi)v_{r'}(q)\overline{v}_{s'}(q)\overline{E}_{q}(\vphi')\gamma^{5}E_{p}(\vphi')u_{s}(p).
\]
Since the electromagnetic background is of a finite extent (disappears at the boundaries of integration), it is more useful to consider probabilities than cross-sections (which would be spacetime-dependent as they depend on the background field strength).  The decay probability $\tsf{P}_{\ps}$ (we use $\tsf{P}_{\ps}$ for pseudoscalar decay and $\tsf{P}_{\sca}$ for scalar decay to reflect their behaviour under parity transformation) is defined as 
$\tsf{P}_{\ps} = V^{2} \int \sum_{\tsf{spin}}\tr|\tsf{S}_{\tsf{fi}}|^{2}\, d^{3}p\,d^{3}q/(2\pi)^{6}$, leading to
\bea
\tsf{P}_{\ps} &=& \frac{g^{2}_{\phi e} \,\delta^{\perp,-}\left(0\right)}{8(\vkap^{0})^{2}k^{0}V} \int \frac{d^{2}p^{\perp}\,dp^{-}}{q^{-}p^{-}} ~\theta\left(p^{-}\right)\theta\left(q^{-}\right)\nn \\
&& \int d\sigma\,d\theta~\tsf{T}\, \exp \left[i\int_{\sigma-\theta/2}^{\sigma+\theta/2}\left[\frac{p\cdot a (\phi)}{\vkap \cdot p }-\frac{q\cdot a(\phi)}{\vkap \cdot q} - \frac{a^{2}(\phi)}{2} \frac{\vkap \cdot k}{\vkap \cdot p ~ \vkap \cdot q}\right]d\phi + ir\theta \right],\nn\\ \label{eqn:P1}
\eea
where momentum conservation is enforced via $q^{\perp,-} = k^{\perp,-} - p^{\perp,-}$ together with the on-shell condition $q^{2}=m^{2}$.  In preparation for eventually performing a local expansion, we have defined the average and difference phase variables
\bea
\sigma = \dfrac{\vphi+\vphi'}{2}; \quad \theta= \vphi-\vphi'. \label{eqn:sigdef}
\eea
The probability can be written in a much neater way by observing that
\bea
r = \frac{k\cdot p}{\vkap \cdot q} - \frac{m_{\phi}^{2}}{2\vkap\cdot q}.
\eea 
We also note
\[
 \delta^{\perp,-}(0) = \lim_{l\to 0} \frac{\delta^{\perp,-}(l)\delta^{+}(l)}{\delta^{+}(l)} = \frac{1}{(2\pi)^{3}} \frac{V\int dt}{\int dx^{-} }  = \frac{Vk^{0}}{(2\pi)^{3}k^{-}} \label{eqn:delta0}
\]
(where we note that $d\vphi/d\tau$ is a constant in a plane wave, where $\tau$ is the proper time, allowing us to cancel the integrals and introduce corresponding momentum factors). 

Then we have
\bea
\tsf{P}_{\ps} &=& \frac{g^{2}_{\phi e}}{4(\vkap^{0})^{2}k^{0}(2\pi)^{3}} \int \frac{d^{2}p^{\perp}\,dp^{-}}{q^{-}p^{-}} ~\theta\left(p^{-}\right)\theta\left(q^{-}\right)\int d\sigma\,d\theta~\tsf{T}\, \mbox{e}^{i\theta\big[\big\langle\frac{k\cdot \Pi}{\vkap\cdot q}\big\rangle - \frac{\mphi^{2}}{2\vkap\cdot q}\big]}, 
\eea
where the classical plane-wave momentum of the electron is
\bea
\Pi = p - a + \vkap ~\frac{2a\cdot p - a^{2}}{2\vkap \cdot p}
\eea
and we define the phase-window-average
\bea
\langle f \rangle \defto \frac{1}{\theta}\int^{\sigma+\theta/2}_{\sigma-\theta/2} f(\phi) d\phi. \label{eqn:wa}
\eea
Performing the spin-sum and the trace we find
\[
\frac{\tsf{T}}{4} = m^{2} + p\cdot q + \frac{\left[a(\vphi)+a(\vphi')\right]\cdot p}{2}\, \frac{\vkap \cdot k}{\vkap \cdot p} - \frac{\left[a(\vphi)+a(\vphi')\right]\cdot q}{2}\, \frac{\vkap \cdot k}{\vkap \cdot q} - \frac{a(\vphi)\cdot a(\vphi')}{2} \frac{(\vkap \cdot k)^{2}}{\vkap \cdot p~\vkap \cdot q}. \label{eqn:T1}
\]

To proceed, we wish to perform the $p^{\perp}$ integrals. We note from \eqnref{eqn:P1} that the exponent is of the form of a Gaussian oscillation in these variables, but the pre-exponent in \eqnref{eqn:T1} also contains terms quadratic in $p^{\perp}$ (in $p\cdot q$). This would seem to lead to a divergence, however, we will show that the divergent contribution can be reinterpreted as an integral over surface terms, which must disappear.
\newline

First of all, we can remove explicit dependence on $q$ (which only remains in the $p\cdot q$ term) by using the trick:
\[
k+ \lambda \vkap = p + q, \qquad \lambda = \frac{2 k\cdot p - m_{\phi}^{2}}{2\vkap \cdot q},
\]
where $\lambda$ was found from the first equation using the on-shell condition $q^{2}=m^{2}$. Then making the replacement
\[
m^{2} + p \cdot q = k\cdot p \frac{\vkap\!\cdot\!k}{\vkap\!\cdot\!q} - m_{\phi}^{2} \frac{\vkap\!\cdot\! p}{2\vkap\!\cdot\! q},
\]
the pre-exponent starts to look like the exponent. Writing this explicitly as
\[
\exp\left[i(\ldots)\right] = \exp\left\{i\left[\theta\left(\frac{k\!\cdot\!p}{\vkap\!\cdot\!q} - \frac{m_{\phi}^{2}}{2\vkap\!\cdot\!q}\right)+\int_{\sigma-\frac{\theta}{2}}^{\sigma+\frac{\theta}{2}}-\frac{k\!\cdot\!a(\phi)}{\vkap\!\cdot\!q}+\frac{\vkap\!\cdot\!k}{2\vkap\!\cdot\!p}(2a(\phi)\!\cdot\!p-a^{2}(\phi)) ~d\phi\right]\right\},
\]
we can make the following replacement in the pre-exponent
\bea
\vkap \!\cdot\! k \frac{k\!\cdot\!p}{\vkap \!\cdot\! q} &\to& -i \vkap \!\cdot\! k \partial_{\theta}+\frac{\vkap \!\cdot\!k}{2\vkap \!\cdot\!q }m_{\phi}^{2} + \frac{\vkap \!\cdot\!k}{\vkap \!\cdot\! q} \frac{k\!\cdot\!(a(\phi)+a(\phi'))}{2}\nn \\
&& +\frac{(\vkap \!\cdot\!k)^{2}}{2\vkap\!\cdot\!q \vkap \!\cdot\!p}\left[-p\cdot(a(\phi)+a(\phi'))+\frac{a^{2}(\phi)+a^{2}(\phi')}{2}\right],
\eea
which simplifies the pre-exponent considerably such that, in the end, we just have
\[
\frac{\tsf{T}}{4} \e^{i(\ldots)}= \left[\frac{m_{\phi}^{2}}{2} + \frac{(\vkap \!\cdot\!k)^{2}}{2\vkap \!\cdot\!q \vkap \!\cdot\!p}\frac{\left[a(\phi)-a(\phi')\right]^{2}}{2} - i \vkap \!\cdot\! k\,\partial_{\theta}\right]\e^{i(\ldots)}.
\]
If we assume that there can be no contribution to the probability in the infinite past or infinite future, we can discard the derivative term in the pre-exponent and perform the $p^{\perp}$ integrals without encountering a divergence.  
\newline

Let us write probabilities in the following way:
\bea
\tsf{P} = \frac{g^{2}}{4\pi} \frac{1}{\eta_{k}} \mathcal{I}, \label{eqn:Pform}
\eea
where the coupling and flux prefactors have been separated from the process-dependent integration, $\mathcal{I}$. Then we define the probability $\tsf{P}_{\ps}$ for the decay of a pseudoscalar into an electron-positron pair as $\tsf{P}_{\ps} = (g^{2}/4\pi\eta_{k})\mathcal{I}_{\ps}$. Selecting a linearly-polarised background, we then eventually arrive at
\bea
\mathcal{I}_{\ps} = \frac{i}{4\pi}\int d\sigma\,dt\, \frac{d\theta}{\theta+i\eps} \left\{\delta^{2} +\frac{\left[a(\phi)-a(\phi')\right]^{2}}{2t(1-t)}\right\}\e^{\frac{i\theta\mu(\theta)}{2\eta_{k}t(1-t)}-\frac{i\theta\delta^{2}}{2\eta_{k}}}\label{eqn:Ips}
\eea
where we define the Kibble mass factor
\bea
\mu(\theta) = 1 + \Big\langle \frac{a}{m}\Big\rangle^{2} - \Big\langle\left(\frac{a}{m}\right)^{2}\Big\rangle, \label{eqn:Km}
\eea
lightfront momentum fraction $t=p^{-}/k^{-}$, and the energy parameter $\eta_{k} = \vkap \cdot k / m^{2}$. (The energy parameter can be thought of as the squared ratio of the centre-of-mass energy to the pair rest energy, for the case when the seed photon collides with a single background photon to produce a pair.) It is possible to perform the $t$-integral analytically (see e.g. \cite{dinu14a})
\bea
\mathcal{I}_{\ps} = \frac{1}{8\pi} \int d\sigma\, \frac{d\theta}{\theta+i\eps} \left\{h(\theta)\delta^{2}
\K_1\left[ih(\theta) \right] +\left[h(\theta)\delta^{2}+i\left[a(\phi)-a(\phi')\right]^{2}\right]\K_0\left[ih(\theta)\right]
\right\}\e^{-ih(\theta)-\frac{i\theta \delta^{2}}{2\eta_{k}}},
\label{eqn:PpsBessel} \nn \\
\eea
where $h(\theta) = -\theta\mu(\theta)/2\eta_{k}$ and $\K_n(x)$ is the modified Bessel function of second kind \Ben{\cite{olver97}}. However \eqnref{eqn:Ips} will prove to be the more useful form of the probability for numerical evaluation.
\newline

Without further derivation, in the spirit of \eqnref{eqn:Pform}, we state that the probability $\tsf{P}_{\sca}$ for the decay of a scalar into an electron-positron pair is proportional to the integral:
\bea
\mathcal{I}_{\sca} = \frac{i}{4\pi}\int d\sigma\,dt\, \frac{d\theta}{\theta+i\eps} \left\{4-\delta^{2} +\frac{\left[a(\phi)-a(\phi')\right]^{2}}{2t(1-t)}\right\}\e^{\frac{i\theta\mu(\theta)}{2\eta_{k}t(1-t)}-\frac{i\theta\delta^{2}}{2\eta_{k}}} ,\label{eqn:Ps}
\eea
and the probability $\tsf{P}_{\gamma}$ for the decay of an unpolarised massive \tsf{U(1)} boson is proportional to the integral:
\bea
\mathcal{I}_{\gamma} = \frac{i}{4\pi} \int d\sigma\,dt\, \frac{d\theta}{\theta+i\eps} \left\{2\left(1-\frac{\delta^{2}}{2(1-t)}\right) - \left[a(\vphi)-a(\vphi')\right]^{2}\left(1-\frac{1}{2t(1-t)}\right) \right\}\e^{\frac{i\theta\mu(\theta)}{2\eta_{k}t(1-t)}-\frac{i\theta\delta^{2}}{2\eta_{k}}}.\label{eqn:Pg} \nn \\
\eea
%
%
%
%
%
%
%
\section{Constant Fields} \label{constantfields}
The Locally Constant Field Approximation (LCFA) allows one to calculate probabilities for processes in non-trivial plane-wave electromagnetic backgrounds by performing a local field expansion of the background, and integrating the resulting constant field result over the non-trivial form of the plane-wave. It has been shown to be a good approximation \cite{nikishov64,king15d} when the intensity parameter of the background $\xi$, satisfies $\xi \gg 1$, where $\xi^{2} = \langle p \cdot T(\vphi) \cdot p \rangle_{\vphi} / m^{2}\,(\vkap\cdot p)^{2}$, for massive seed particle four-momentum $p$, stress-energy tensor $T^{\mu\nu} = (F^{2})^{\mu\nu}-\eta^{\mu\nu}\tr F^{2}/4$, $F$ is the Faraday tensor and $\langle\cdot\rangle_{\vphi}$ implies a cycle-average over the phase $\vphi$ \cite{ilderton09}. (However, recent analyses of nonlinear Compton scattering hint that the infra-red behaviour is badly approximated by the LCFA \cite{dipiazza17z,king18d}.) (The LCFA is sometimes explained by reference to when a massive seed particle is highly relativistic, the electromagnetic field in the particle's rest-frame is approximately that of a constant-crossed field \cite{ritus85}.)
\newline

One can acquire the LCFA result from the probability for a process in a plane-wave pulse, such as \eqnref{eqn:Ips}, by expanding the exponent in $\theta$ up to $\mathcal{O}(\theta^3)$ which corresponds to the highest power contributing to the constant field case. This amounts to making the replacements
\bea
\theta\mu &\to& \theta + \frac{f^{2}(\sigma)}{12}\theta^{3}\qquad 
\left(\mbf{a}(\phi)-\mbf{a}(\phi')\right)^{2} \to -\theta^{2}f^{2}(\sigma),
\eea
where $f(\sigma) = \xi'(\sigma)$ and the linearly-polarised background can be written $a = -\eps (a\cdot \eps)$ and $-\eps\cdot a'(\sigma) = mf(\sigma)$.
Using the results:
\bea
\int_{-\infty}^{\infty}\frac{d \theta}{\theta+i\eps} \e^{i(r\theta + c_{3}\theta^{3})} 
= - 2\pi i \Ai_{1}\left[\frac{r}{(3c_{3})^{1/3}}\right]; \quad
\int_{-\infty}^{\infty}d \theta~\theta \e^{i(r\theta + c_{3}\theta^{3})} = -\frac{2\pi i}{(3c_{3})^{2/3}}\,\Ai'\left[\frac{r}{(3c_{3})^{1/3}}\right],\nn \\
\eea
for $c_{3} \in \mathbb{R}$, we then find
\bea
\mathcal{I}^{\tsf{LCFA}}_{\ps} =  \int d\sigma\,dt \left\{\frac{\delta^{2}}{2}\,\Ai_{1}(z) - \chi_{k}(\sigma)\sqrt{z_{0}}\,\Ai'(z)\right\} \label{eqn:Pe2},
\eea
where we define
\bea
z_{0} = \left(\frac{\chi_{k}(\sigma)}{\chi_{p}(\sigma)\chi_{q}(\sigma)}\right)^{2/3}; \quad z = z_{0} - \frac{\delta^{2}}{\chi_{k}(\sigma)\sqrt{z_{0}}},
\eea
$\chi_{k}(\sigma) = f(\sigma)\eta_{k}$. (Applying the above procedure to the \tsf{U(1)} case \eqnref{eqn:Pg} and taking the massless limit $\delta^{2}\to0$, leads exactly to the QED expression for photon-seeded pair-creation \cite{elkina11}. Furthermore the mass-dependent part of the Airy argument has the same form as  expected from e.g. the second step of electron-seeded pair-creation, given explicitly in \cite{king13b}.)
\newline

Typically, we are interested in lab-based detection of ALPs using constant \emph{magnetic} fields. Clearly, a constant magnetic field, which can be written $a(\vphi) \sim \vphi$ for spacelike ($\vkap^{2}<0$) wavevector, is relativistically inequivalent to a constant crossed field (CCF), which has equal magnitude electric and magnetic fields. However, if we restrict our analysis to highly relativistic seed particles, i.e. with $k^{-}/m \gg 1$, then from their rest frame, a constant magnetic field will appear to be well-approximated by a CCF \cite{jackson99}. This fact, which underlies the Weizs\"{a}cker-Williams approximation \cite{weizsaecker34,williams34}, was recently explicitly shown to hold for nonlinear Thomson scattering in a constant magnetic field \cite{king16b}. The constant-field limit $a(\vphi) \sim \vphi$ for lightlike wavevector is then the constant crossed field limit.
\newline

To approximate the probability in a constant magnetic field using a CCF, we can relate the scaled vector potential, $a$, to the field-strength $\F$ by first writing $\F$ as
\[
\F(\vphi) = \partial_{t}A^{1}(\vphi) = \frac{1}{e}\frac{\partial\vphi}{\partial t}a^{1\,\prime}(\vphi),
\]
where we pick the background to be polarised in the $1$-direction without loss of generality. To proceed in evaluating $\partial \vphi/\partial t$, we use the same reasoning as in \eqnref{eqn:delta0}, then we see that the non-trivial component of the reduced vector potential, $a^{1}(\vphi)$, can be written
\bea
a^{1}(\vphi) = m\xi(\vphi) \to m\frac{k^{0}}{k^{-}}\frac{L}{\lambda_{\tsf{C}}}\frac{\F_{0}}{\Fs}\,\vphi, \label{eqn:CCFa}
\eea
where we have introduced a nominal ``frequency'' of the constant field $\vkap^{0}$ to be $\vkap^{0} = 2\pi/L$ with $L$ being the longitudinal spatial extent of the constant field (formally infinite in the CCF limit), $\lambda_{\tsf{C}} =2\pi/m$ is the Compton wavelength, $\F_{0}$ is the amplitude of the field strength, and $\Fs = m^{2}/e$ is the Schwinger limit. Other quantities can then be written independent of any external field frequency:
\[
\chi_{k}(\sigma) \to \chi_{k} = f\eta_{k} = \frac{F}{\Fs}\frac{k^{-}}{m}; \qquad \frac{1}{\eta_{k}}\int d\sigma \to \frac{m}{k^{-}}\frac{L}{\lambdabar}.
\]
Then we see:
\bea
\tsf{P}^{\tsf{\tiny{CCF}}}_{\ps} = \frac{g^{2}}{4\pi}\frac{m}{k^{-}}\frac{L}{\lambdabar}\,\tsf{R}_{\ps}^{\tsf{\tiny{CCF}}}(\chi_{k}, \delta^{2}),
\eea
where all the non-trivial dependency on experimental parameters is contained within the function  $\tsf{R}_{\ps}^{\tsf{\tiny{CCF}}}$, which is the rate per unit detector length (measured in units of the reduced Compton wavelength).
\newline

In relation to the ALP mass and the field strength we can identify three distinct regimes for creation of electron-positron pairs: i) below threshold, $\delta^{2} <4$, where the process is forbidden in the limit of zero field and hence is \emph{field-induced}; ii) above threshold, $\delta^{2}>4$, where the process is \emph{field-assisted} and can proceed in the zero-field limit, and iii) strong field $\chi\gg 1$, where decay is so likely, there is no threshold behaviour anymore. Plotting the dependency of $\tsf{R}_{\ps}^{\tsf{\tiny{CCF}}}$ on $\chi_{k}$ in \figref{fig:del2a} for different axion mass parameter, $\delta^{2}$, one can clearly see where these three regions occur.
\newline

\begin{tikzpicture}[domain=0:4] 
\node at (0,0) {\includegraphics[width=7cm]{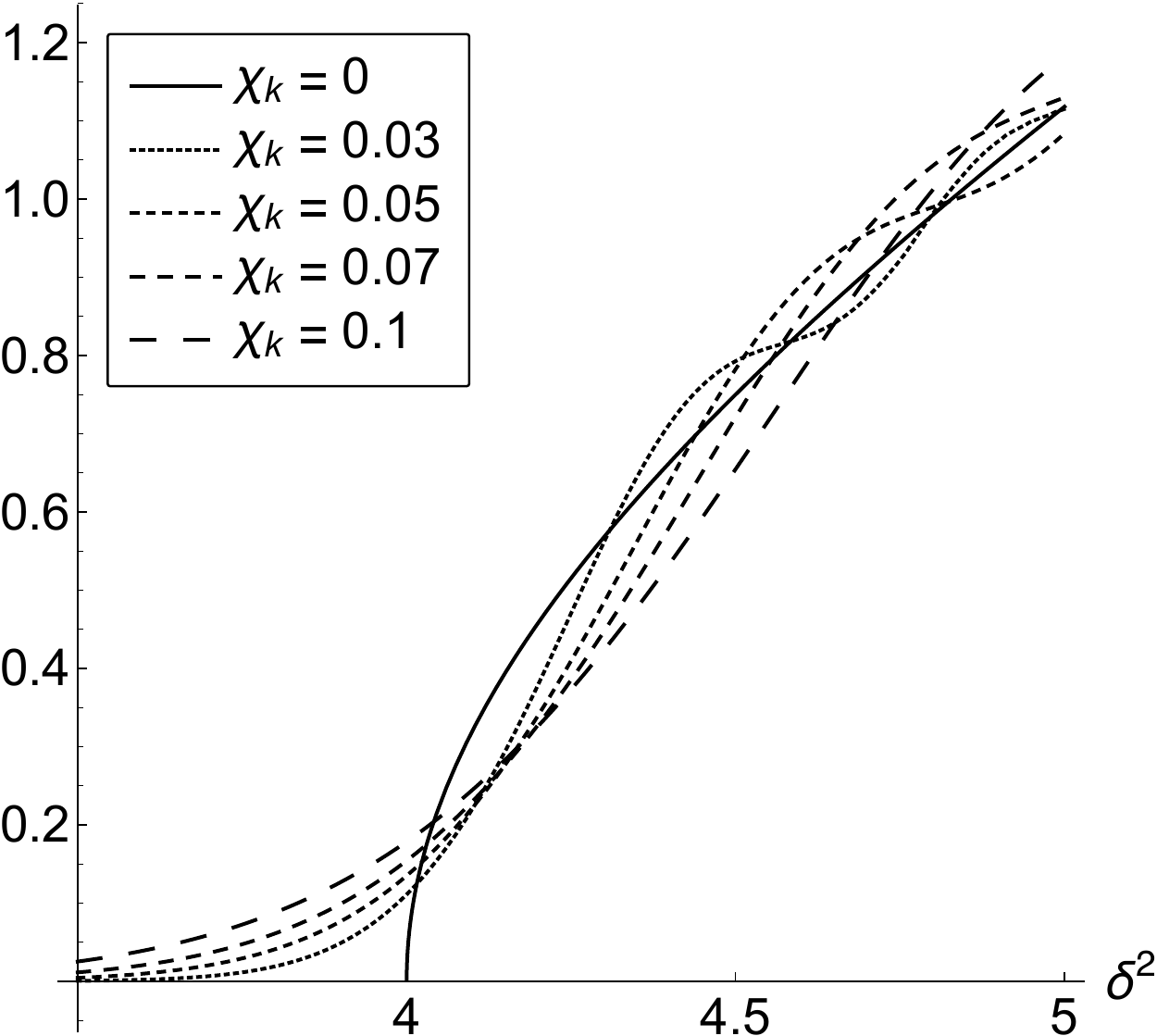}};
\node[draw] at (-2.6,3.5) {a};
\node at (8,0) {\includegraphics[width=7cm]{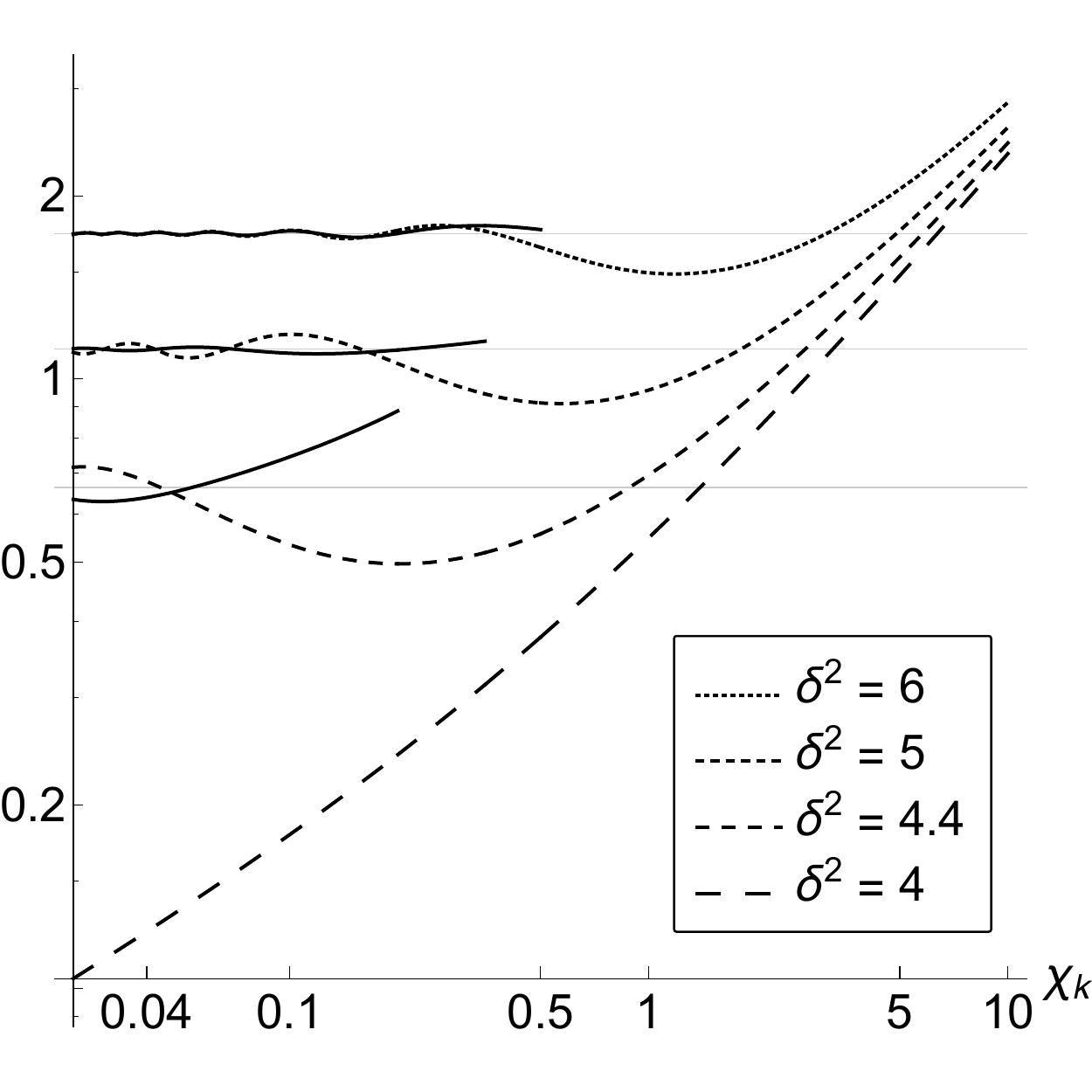}};
\node[draw] at (5.3,3.5) {b};
\end{tikzpicture}
\begin{figure}[h!!]
\centering
\caption{Plots of $\tsf{R}_{\ps}^{\tsf{\tiny{CCF}}}$ for various seed-particle masses. a) the effect of the threshold when $\chi_{k}\ll1$ -- exponential suppression below the threshold and an oscillatory dependence on $\delta^{2}$ above it. b) for small $\chi_{k}$, when the decay is above-threshold, the probability oscillates about the vacuum values (shown as grey horizontal lines) and in the below-threshold regime, the probability is exponentially suppressed. For large $\chi_{k}$, the distinction between above- and below- threshold is lost and the dependence on particle mass disappears. In b), black solid lines are the asymptotic results, which become more accurate for heavier masses and smaller $\chi_{k}$.}\label{fig:del2a}
\end{figure}

\subsection{Below threshold decay}
Below-threshold, pair creation can only occur as a tunneling process and hence is exponentially suppressed. This situation is very similar to studies on photon-seeded pair-creation \cite{reiss62,nikishov64}, and so we will not analyse it in great detail. However, we highlight the difference that a massive seed particle brings, by calculating the asymptotic and perturbative limits of $\tsf{R}_{\ps}^{\tsf{\tiny{CCF}}}$. 
\newline

In a CCF, the exponent in \eqnref{eqn:Ips} (and \eqnrefs{eqn:Ps}{eqn:Pg}), can be written, with a change of variable to make the dependency on $\chi_{k}$ manifest,  $\theta \to 2\vartheta/\xi$, as:
\bea
\exp\left[\frac{i\vartheta}{\chi t(1-t)} \left( 1 + \frac{\vartheta^{2}}{3}- \delta^{2}t(1-t)\right)\right]. \label{eqn:exp1}
\eea
In the subthreshold case, $\delta^{2}<4$, we note that $1-\delta^{2}t(1-t)>0$. Then in this case, the turning points of the exponent always lie on the imaginary axis. Rotating the integration contour with $ \vartheta \to i \vartheta$,
yields a real exponent of the form $\sim \vartheta - c \vartheta^{3}$, allowing us to use the Laplace method. There are two turning points and one is subdominant.
\newline

We then find for $\delta^{2}<4$ and $\chi_{k} \ll 1$:
\bea
\tsf{R}_{\ps}^{\tsf{\tiny{CCF}}} &\sim& \chi_{k}\frac{\sqrt{3}}{4\sqrt{2}} \left(1+ \frac{\delta^{2}}{8}\right)^{-1/2}\left(1+ \frac{\chi_{k}}{2} \frac{\delta^{2}}{4-\delta^{2}}\right)\e^{-\frac{8}{3\chi_{k}}\left(1-\frac{\delta^{2}}{4}\right)^{3/2}}. 
\label{eqn:BelowThresholdIntegral}
\eea
We notice the familiar $-8/3\chi_{k}$ tunneling exponent has been shifted by the seed particle mass, where the tunneling behaviour clearly disappears as the mass approaches the threshold $\delta^{2}\to 4$. We can understand this in an intuitive way by using arguments based on energy-momentum conservation \cite{akhmedov11}, recently applied to high-intensity laser-based QED \cite{fedotov15}. The energy of the produced electron is:
\bea
\mathcal{E}_{p}(t) = \sqrt{\left(\mbf{p}-e\mbf{A}\right)^{2}+m^{2}} = \sqrt{p^{2}+m^{2} + e^{2}F^{2}t^{2}},
\eea
where in the last equality, we have used the fact that the background field is constant. Assuming all particles involved in the decay are highly relativistic, we can see that the energy change is:
\bea
\Delta \mathcal{E}(t) = \mathcal{E}_{p}(t) + \mathcal{E}_{k-p}(t)-\mathcal{E}_{k} \approx \frac{2m^{2}}{k} \left\{\left[1+\frac{(eFt)^{2}}{m^{2}}\right] \frac{k^{2}}{4p(k-p)} - \frac{\delta^{2}}{4}\right\}. \label{eqn:DelE1}
\eea
We can approximate the form of the rate for the process to occur using the WKB method \cite{bender78}:
\bea
\frac{d\tsf{P}}{dt} \sim \exp\left[ i \int_{0}^{t} \Delta \mathcal{E}(t') dt'\right] \sim \tsf{R}_{\ps}^{\tsf{\tiny{CCF}}}. \label{eqn:wkb1}
\eea
For a tunneling process, we can approximate this integral by using the saddle-point method, and integrating to $t_{\ast}$ where $t=t_{\ast}$ is the shortest time for which $\Delta \mathcal{E}(it) = 0$. The smallest energy difference corresponds to an equal distribution of the initial energy and momentum $p=k/2$, from which it follows that the tunneling time is:
\[
t_{\ast} = \frac{m}{eF}\,\sqrt{1- \frac{\delta^{2}}{4}},
\]
and using \eqnref{eqn:wkb1}, we indeed find that:
\[
\tsf{R}_{\ps}^{\tsf{\tiny{CCF}}} \sim \exp \left[-\frac{8}{3\chi_{k}}\left(1-\frac{\delta^{2}}{4}\right)^{3/2}\right].
\]
Therefore, we can be somewhat confident that we have the correct tunneling exponent.

\subsection{Above threshold decay}
Above threshold, $\delta^{2}>4$,a region of the $t$-integration exists where $1-\delta^{2}t(1-t)<0$. Then in this case, two turning points of the exponent  \eqnref{eqn:exp1} appear, with opposite sign, on the real $\vartheta$ axis. 

Then applying the method of stationary-phase in $\vartheta$, we acquire a final integral in $t$, with an oscillating exponent which also has turning points on the real axis. Of those, two turning points conspire to produce a cosine, and the third turning point gives a constant term, which is where the zero-field contribution originates. Altogether we find when $\chi_{k}\ll1$, $\delta^{2}>4$, that:
\bea
\tsf{R}_{\ps}^{\tsf{\tiny{CCF}}} &\sim& -\frac{\chi_k\sqrt{3}}{8} \left(1+ \frac{\delta^{2}}{8}\right)^{-1/2}\left(4 + \frac{\delta^{2}}{\delta^{2}-4}\right)\cos\left[\frac{8}{3\chi_{k}}\left(-1+\frac{\delta^{2}}{4}\right)^{3/2}\right]+\frac{1}{2}\sqrt{\delta^{2}\left(\delta^{2}-4\right)}. \nn \\ \label{eqn:asyAT}
\eea

We see that in this low-$\chi_{k}$ limit, the probability for decay oscillates as the field-strength is varied. This is demonstrated for various axion masses in \figrefa{fig:del2a}, where we plot this transition, and the dependency of $\tsf{R}_{\ps}^{\tsf{\tiny{CCF}}}$ on ALP mass parameter, $\delta^{2}$, where the threshold effect can clearly be seen.

The zero-field result in \eqnref{eqn:asyAT} must, of course, be independent of the form of the background field. By taking the limit $a\to 0$ in \eqnref{eqn:Ips}, we find the same result, which can be written as:
\[
\tsf{P}_{\ps}(\xi\to0) \to \frac{g^{2}}{4\pi} \frac{m T}{2}\sqrt{\delta^{2}(\delta^{2}-4)},
\]
where $T = \int dt$.

\subsection{Strong fields, \texorpdfstring{$\chi_{k}\gg1$}{chi much greater than 1}}
In this parameter region, one can simply perturbatively expand \eqnref{eqn:Pe2} in the small parameter $1/\chi_{k}$ since the Airy argument is given by:
\[
z = \frac{1}{\chi_{k}^{2/3}}\left(\frac{1}{t(1-t)}\right)^{2/3}\left[1-\delta^{2}t(1-t)\right].
\]
At a given $\chi_{k} \gg 1$, this perturbative expansion decreases in accuracy for increasing $\delta^{2}$. Suppose $\delta^{2}\gg 1$, then $z \sim \delta^{2}/\chi_{k}^{2/3}$. So in the perturbative limit, we must also assume that $\chi_{k}^{2/3} \gg \delta^{2}$.

After a straightforward integration in $t$, we find:
\bea
\tsf{R}_{\ps}^{\tsf{\tiny{CCF}}}(\chi_{k}, \delta^{2}) &\approx& \frac{2^{4/3}\pi \chi^{2/3}_{k}}{3^{1/3}\Gamma(\nf{1}{6})\Gamma(\nf{7}{6})}+O(\chi_{k}^{0}); \qquad O(\chi_{k}^{0}) = \frac{\delta^{2}}{3} + O(\chi_{k}^{-2/3}).\label{eqn:pertres1}
\eea
We have included the next-to-leading-order term in $\chi_{k}^{2/3}$ to show that, in the limit $\chi_{k}^{2/3} \gg \delta^{2}$, the mass of the seed particle ceases to play a role, and in general, the concept of a threshold disappears as $\chi_{k}$ increases towards $\chi_{k}\gg 1$.

Again, the functional dependence of \eqnref{eqn:pertres1} can be understood by using intuitive methods \cite{akhmedov11,fedotov15}. In this case, the process is above-threshold and so the rate is simply proportional to $g^{2}$ (because one vertex) and $1/t_{q}$ (because of dimensions), where $t_{q}$ is the quantum time fulfilling $t_{q} = 1/\Delta \mathcal{E}(t_{q})$ from the uncertainty relation. (Since the process is quantum, the classical timescale $t_{cl}$ given by $eF(t_{cl})t_{cl} = m$ should not be significant.) Using \eqnref{eqn:DelE1} generates a cubic in $t_{q}$, and the one real root leads to the relation:
\bea
\tsf{R}_{\ps}^{\tsf{\tiny{CCF}}} \sim \chi_{k}^{2/3} + O(\chi_{k}^{0}); \quad O(\chi_{k}^{0}) \sim \left[1-\frac{\delta^{2}}{4}\right] + O(\chi_{k}^{-2/3}).
\eea
Also here, we include the next-to-leading order, to show the dependency on the mass. Although the pseudoscalar mass term in our result \eqnref{eqn:pertres1} depends on the mass as $\sim \delta^{2}$ and not $\sim (1-\delta^{2}/4)$ as in the intuitive method, it is not entirely surprising that this term differs since it originates from the parity-dependent term in the trace, which takes a different form whether dealing with e.g. a photon, scalar or pseudoscalar and nowhere have we inserted the fact that we are dealing with a pseudoscalar in this intuitive picture. (Indeed from \eqnref{eqn:Ps} the scalar mass term is $\sim -(1-\delta^{2}/4)$.)

\section{Edge effects of static constant fields} \label{edgeeffects}

\begin{figure}[h!!]
\centering
\includegraphics[width=7cm]{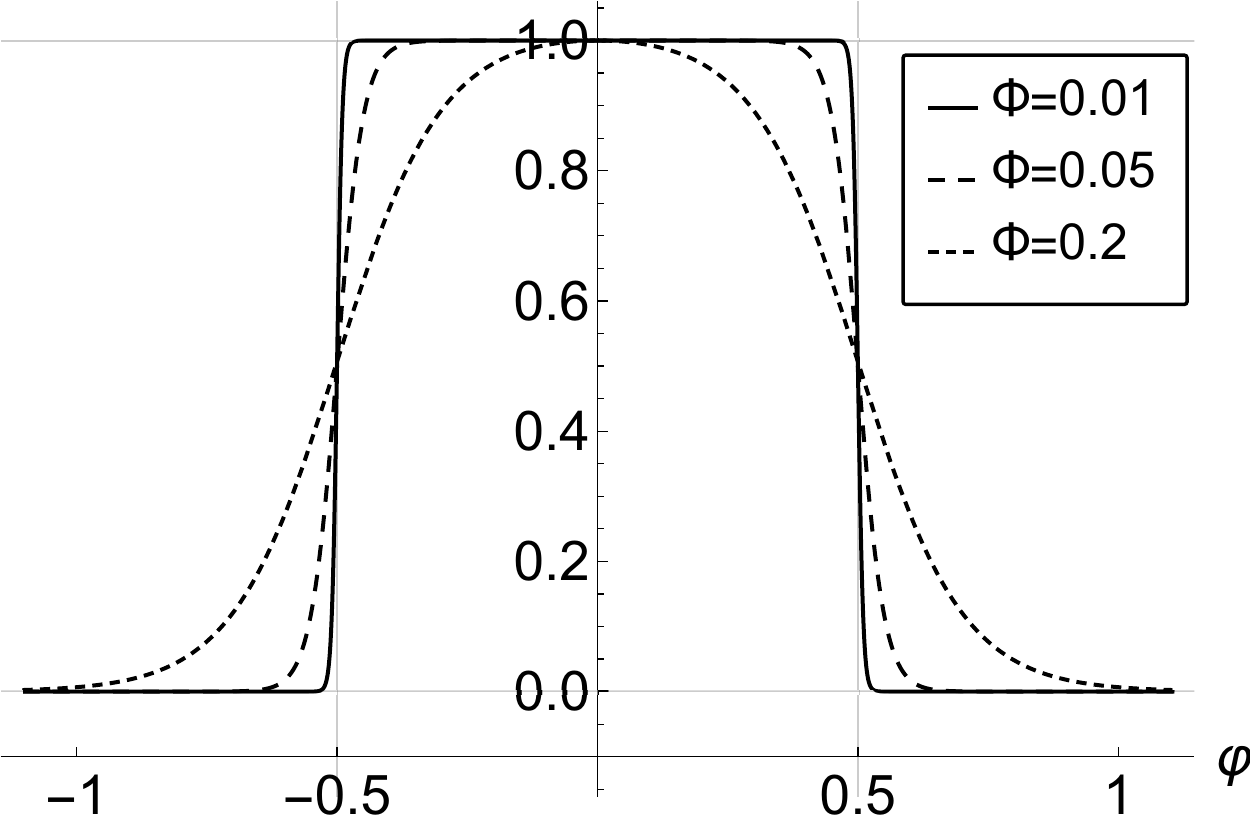}
\caption{Magnetic part of a constant crossed field, scaled so the maximum is unity. For highly-relativistic axions, the crossed field produces an equivalent effect to   entering a constant homogeneous magnetic field. The limit $\Phi\to0$ gives the top-hat function.}\label{fig:Efs1}
\end{figure}

We recall that we are working in the highly-relativistic regime, where processes are well-approximated by replacing a constant field with a CCF. Therefore, to represent the edge of a quasi-constant magnetic field, we choose a field of plane-wave form:
\bea
B(\vphi) = \frac{F_{0}}{1+\tanh(-\vphi_{0}/\Phi)\tanh(\vphi_{1}/\Phi)} \left[1+\tanh\left(\frac{\vphi-\vphi_{0}}{\Phi}\right)\tanh\left(\frac{\vphi_{1}-\vphi}{\Phi}\right)\right], \label{eqn:B1}
\eea
where $\vphi=\vphi_{0,1}$ (and $\vphi_{1}>\vphi_{0}$) are the phase positions of the two ``edges'' of the field and $\Phi$ is a sharpness parameter. We choose, without loss of generality, $\vphi_{0}=-0.5$, $\vphi_{1}=0.5$, and exhibit the form of $B$ for various sharpness parameters in \figref{fig:Efs1}.
The potential is then derived from this form of the field numerically by solving $a'(\vphi) = m (k^{0}/k^{-}) (L/\lambda_{\tsf{C}}) B(\vphi)/\F_{\tsf{Q}}$, which reduces to \eqnref{eqn:CCFa} in the constant-field limit.

To apply the result for the probability for axion decay in a plane-wave field \eqnref{eqn:Ips} to a quasi-static magnetic field in the highly-relativistic regime, we use \eqnref{eqn:CCFa}, whence it follows  $\xi = (L/\lambdabar)(\F_{0}/\Fs)$. For a magnetic field of $\F_{0}=1\,\trm{T}$ and length $L=1\,\trm{m}$, $\xi \sim O(10^{2})$, however the frequency scale $\vkap^{0}/m = 2\pi \lambdabar/ L \ll 1$, so in this case $\eta_{k} \ll 1$ and hence $\chi_{k} = \xi\eta_{k} \ll 1$. Therefore, following from the results in \figref{fig:del2a}, for a terrestrial magnetic field, we expect the pair-decay to only occur for axions with a mass that is close to, or already above threshold. Still, we will begin by analysing the below-the-threshold case as it gives a clearer demonstration of the effect of field gradients introduced by having a rapid drop-off at the end of the magnetic field.
\newline

In this case, the numerical evaluation of \eqnref{eqn:Ips} is non-trivial, because the constant-field part is not absolutely convergent. We found it was sufficient to integrate by parts once in $\theta$ and to numerically evaluate the resulting, absolutely convergent integral.
\newline

Upon comparison with the LCFA, we expect that when $\xi \Phi \not \gg 1$, there may be a discrepancy. 

We can justify the LCFA by applying the substitution $\theta \to y/\xi$ in  \eqnref{eqn:Ips} to give:
\bea
\mathcal{I}_{\ps} = \frac{i}{4\pi} \int d\sigma \,dt\, \frac{dy}{y} \left\{\delta^{2} +\frac{\left[a\left(\sigma+\frac{y}{2\xi}\right)-a\left(\sigma-\frac{y}{2\xi}\right)\right]^{2}}{2t(1-t)}\right\}\e^{\frac{iy}{2\chi_{k}}\left[\frac{\mu(\frac{y}{\xi})}{t(1-t)}-\delta^{2}\right]}.\nn \\
\label{eqn:Ips2}
\eea
Then if $\xi \gg 1$, we expect a Taylor expansion of functions in $y/\xi$ - for example the dimensionless Kibble Mass, $\mu$ -  to be the basis of a good approximation. The conditions that powers of $y$ higher than $y^{3}$ can be discarded - and hence the LCFA used - include such inequalities as $[(y/\xi)(a''(\sigma)/a'(\sigma))]^{2} \ll 1$, $[(y/\xi)^{2}(a'''(\sigma)/a'(\sigma)) \ll 1$, and for these terms to make a difference, the probability must not be already vanishingly small when these inequalities are violated. It then follows that $(\xi \Phi)^{-2}\ll 1$ for the LCFA to be valid. For these parameters, where we have chosen to associate the external-field frequency $\vkap^{0}$ with $2\pi/ L$, if the ALP collides head-on with the wavevector of the magnetic field, $\Phi$ then represents some length or duration, $\Delta$, over which the field falls off at its edges. Then LCFA is valid when:
\bea
\left(\frac{1}{\xi}\frac{L}{\Delta}\right)^{2} \ll 1; \qquad \left(\frac{\lambda_{\tsf{C}}}{\Delta}\frac{\Fs}{\F_{0}}\right)^{2} \ll 1,
\eea
where $\lambda_{\tsf{C}}$ is the Compton wavelength of an electron. Therefore, the weaker the field, the more important its shape. This is somehow intuitive: a weaker field has a lower intensity and so the approximation that it is locally constant should be worse. For a $1\,\trm{T}$ magnet, this corresponds to a field edge of approximately $\Delta^{2} \ll 10^{-4}\,\trm{m}^{2}$. 
\newline

\begin{figure}[h!!]
\centering
\includegraphics[width=5cm]{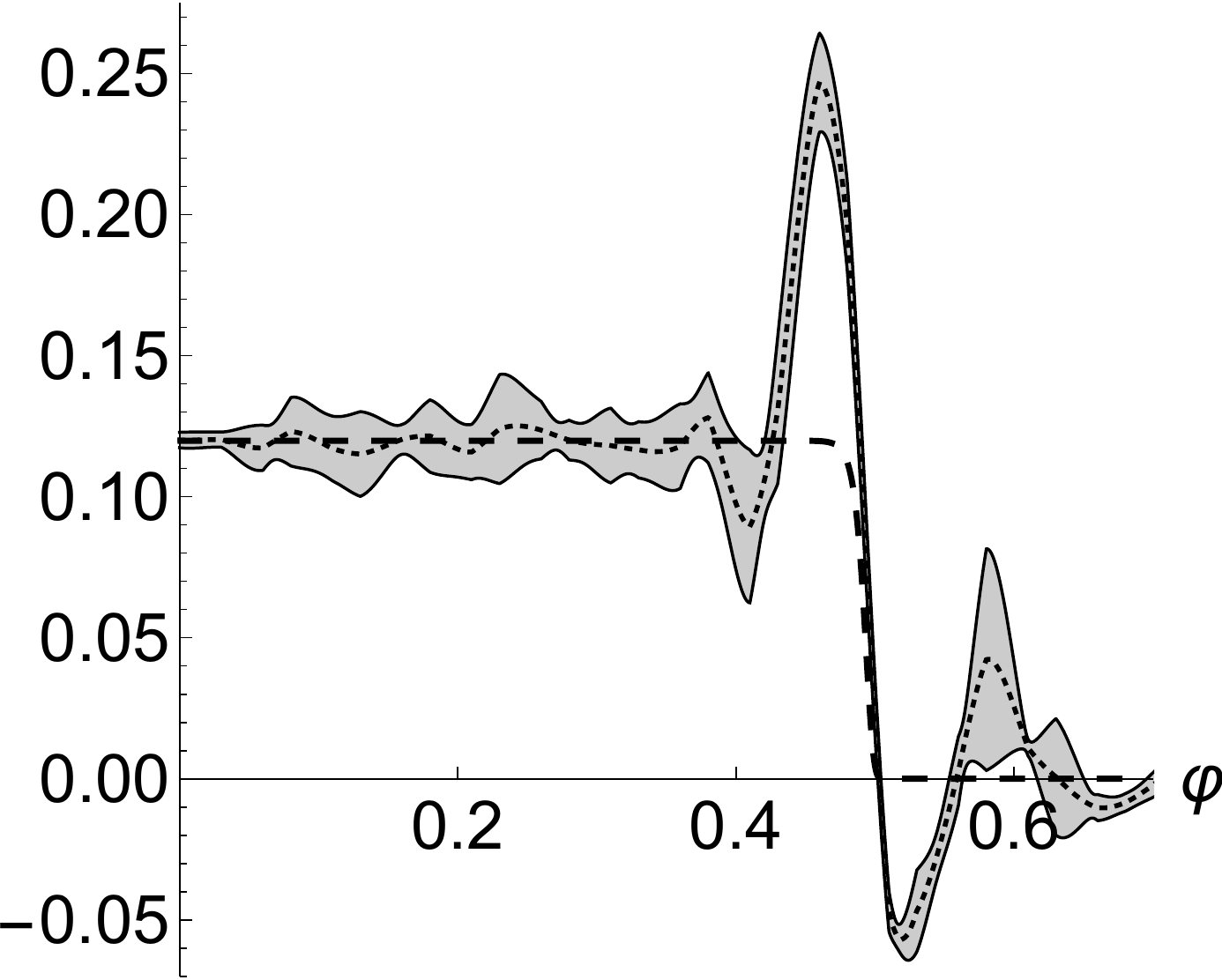}\hfill\includegraphics[width=5cm]{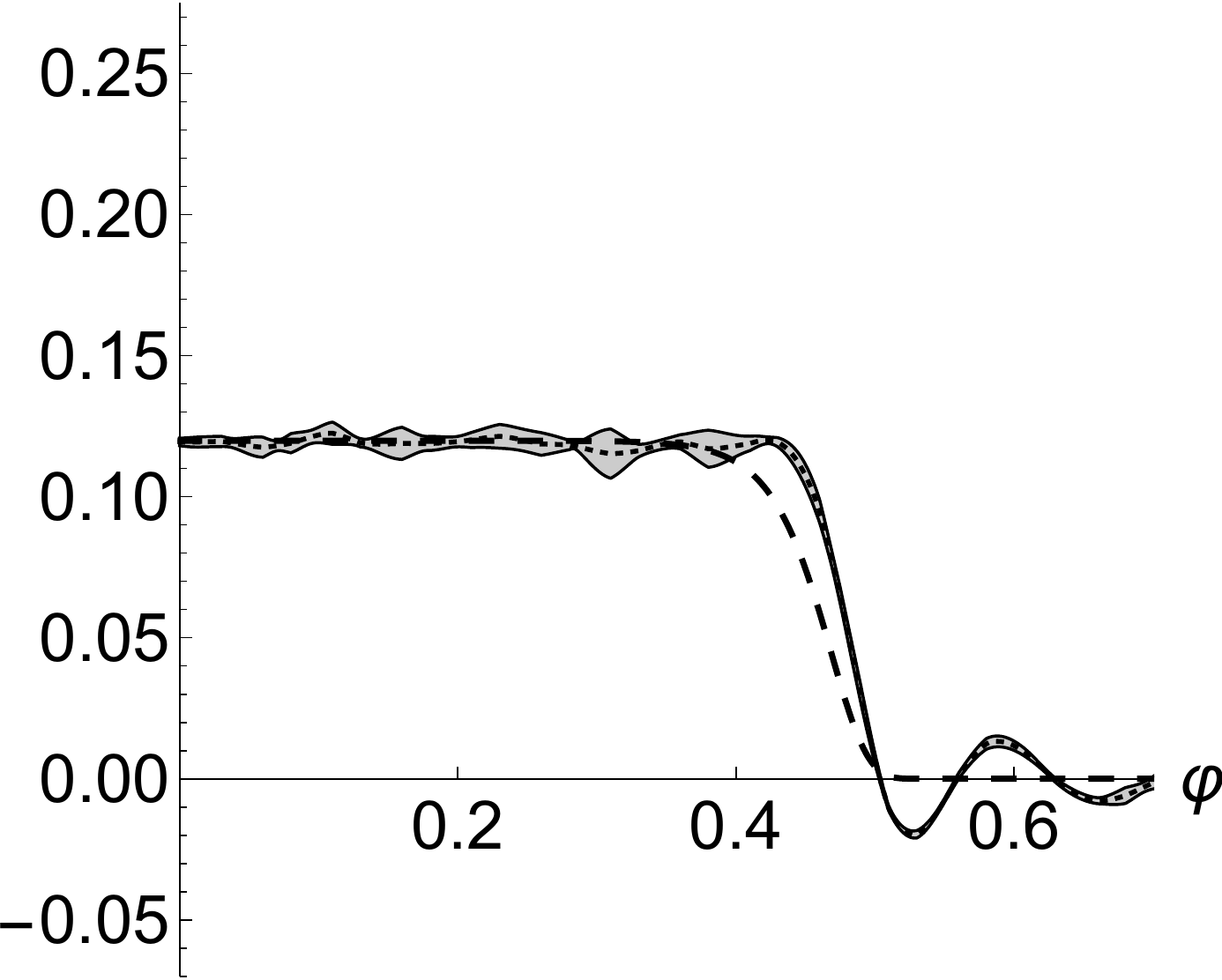}\hfill\includegraphics[width=5cm]{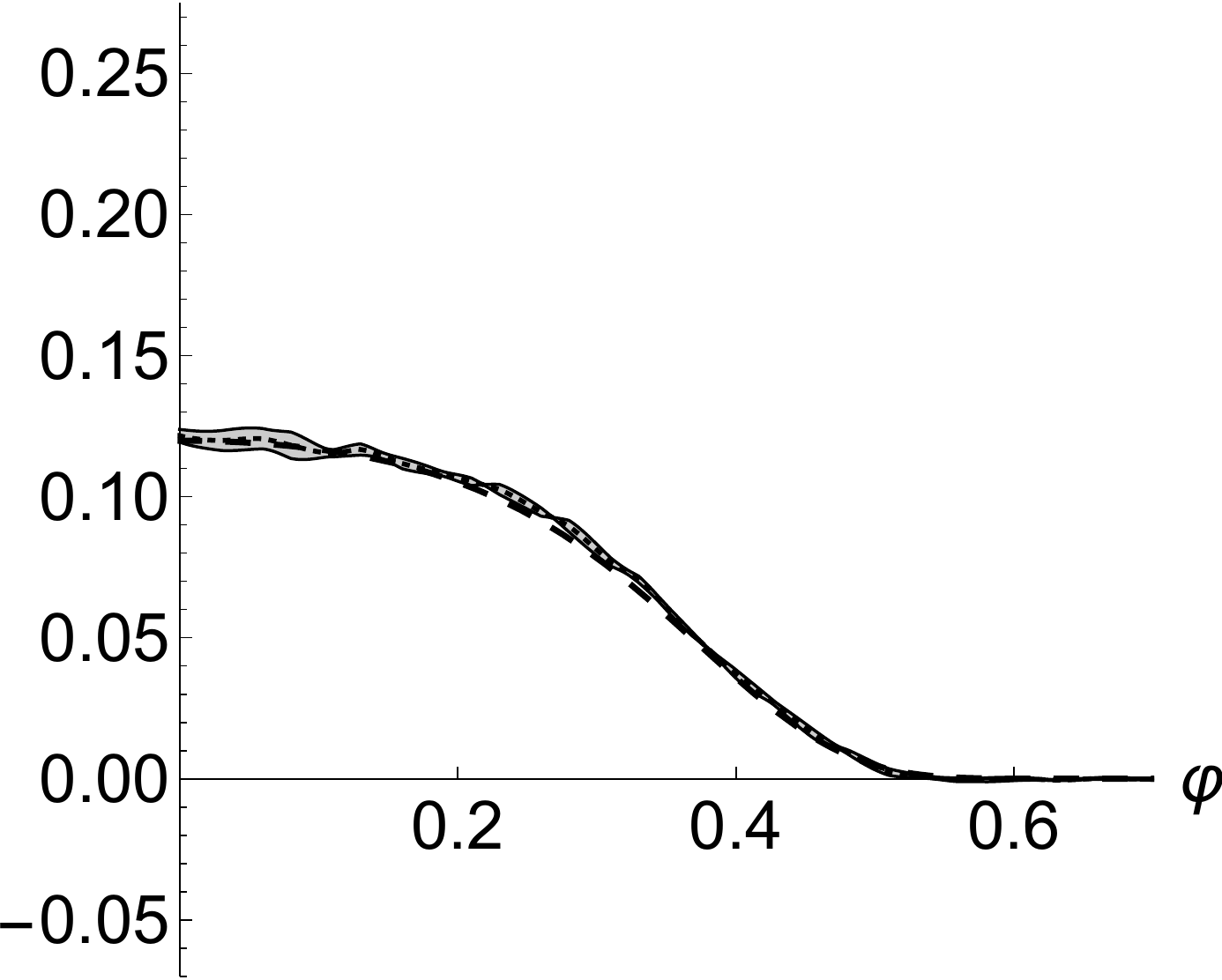}\hfill
\caption{Plots of $\mathcal{I}_{-}$ for a field when $\delta^{2}=0$, $\xi=10$, $\eta_{k} = 0.1$, for three cases: $\Phi=0.2$, $\Phi=0.05$ and $\Phi=0.01$ respectively from left to right. The dashed line is the LCFA, the dotted line is the mean numerical result and the gray area marks out one standard deviation from the mean. The numerics were run five times for each plot, and each numerical evaluation randomly allocated $1000$ points in $\theta$ between $0$ and $1$, which were distributed with a cubic weighting towards $\theta=0$.}\label{fig:edged2}
\end{figure}

We demonstrate the effect of the sharpness of the field in \figref{fig:edged2} for $\xi=10$, $\eta_{k}=0.1$ (a much higher value than from an ALP in a homogeneous magnetic field in the lab, which would require a $100\,\trm{TeV}$ ALP in a $1\,\trm{T}$ magnet), so that $\chi_{k} = 1$ and pair-creation is appreciable.

The numerical curves in \figref{fig:edged2} are also seen to oscillate around zero outside of the magnetic field. It is known \cite{king18d} that $\partial \tsf{P}/\partial \sigma$ (we recall the definition of $\sigma$ in \eqnref{eqn:sigdef} as the seed particle's average phase position) does not have to be positive, as long as the total probability, which is defined for asymptotic states, \emph{is} positive. Therefore in general, $\partial \tsf{P}/\partial \sigma$ cannot be interpreted as a rate. 
 In \figref{fig:edged2}, we note that the introduction of strong field gradients through a sharper magnetic field edge, allows for an increase in the amplitude of $\partial \tsf{P}/\partial \sigma$ near this edge. One way this can be understood is by considering the Fourier transform of the limit $\Phi \to 0$, which would give a frequency spectrum $\sim\trm{sinc}(r\vkap^{0}L)$, where $r$ is a real number. The opposite limit $\Phi \to \infty$ makes the field formally constant resulting in a delta frequency spectrum at the origin. Thus, the higher the field gradient, the larger the contribution from higher frequencies, which can bridge the gap to the $2m$ pair-creation threshold, thereby reducing the necessary tunneling time and increasing the probability. This is similar to the situation of pair-creation by a photon in the background of a plane-wave laser pulse, where shorter pulses were found to drastically increase pair-creation probabilities \cite{titov12,nousch12}. Although homogeneous magnetic field strengths are limited to around $O(1-10)\,\tsf{T}$ in the lab, there is increasing interest in the quasi-static fields of the order of $\sim 10^{5}\,\trm{T}$ that are generated in intense laser-plasma collisions \cite{jansen18}.
\newline

To demonstrate the numerical integration in $\theta$, we 
plot in \figrefa{fig:plotnum} how the integrand in \eqnref{eqn:Ips2} depends on $\theta$, with the same parameters as the central plot in \figref{fig:edged2} ($\delta^{2}=0$, $\Phi=0.05$, $\xi=10$, $\eta_{k}=0.1$) evaluated at $\sigma=0.3$ (as the problem is symmetric around $\theta=0$, we have simply calculated points in the region $\theta\geq0$). The convergence of the integral is indicated in \figrefb{fig:plotnum} where the cumulative distribution function $C(\sigma,\theta)$, given by
\[
 C(\sigma, \theta) = \int_{0}^{\theta} \frac{\partial^{2}\mathcal{I}_{-}(\sigma,y)}{\partial \sigma \partial y} dy,
\]
is plotted and tends towards a constant as the upper integration bound in $\theta$ is increased.
\newline

\begin{tikzpicture}[domain=0:4] 
\node at (0,0) {\includegraphics[width=6cm]{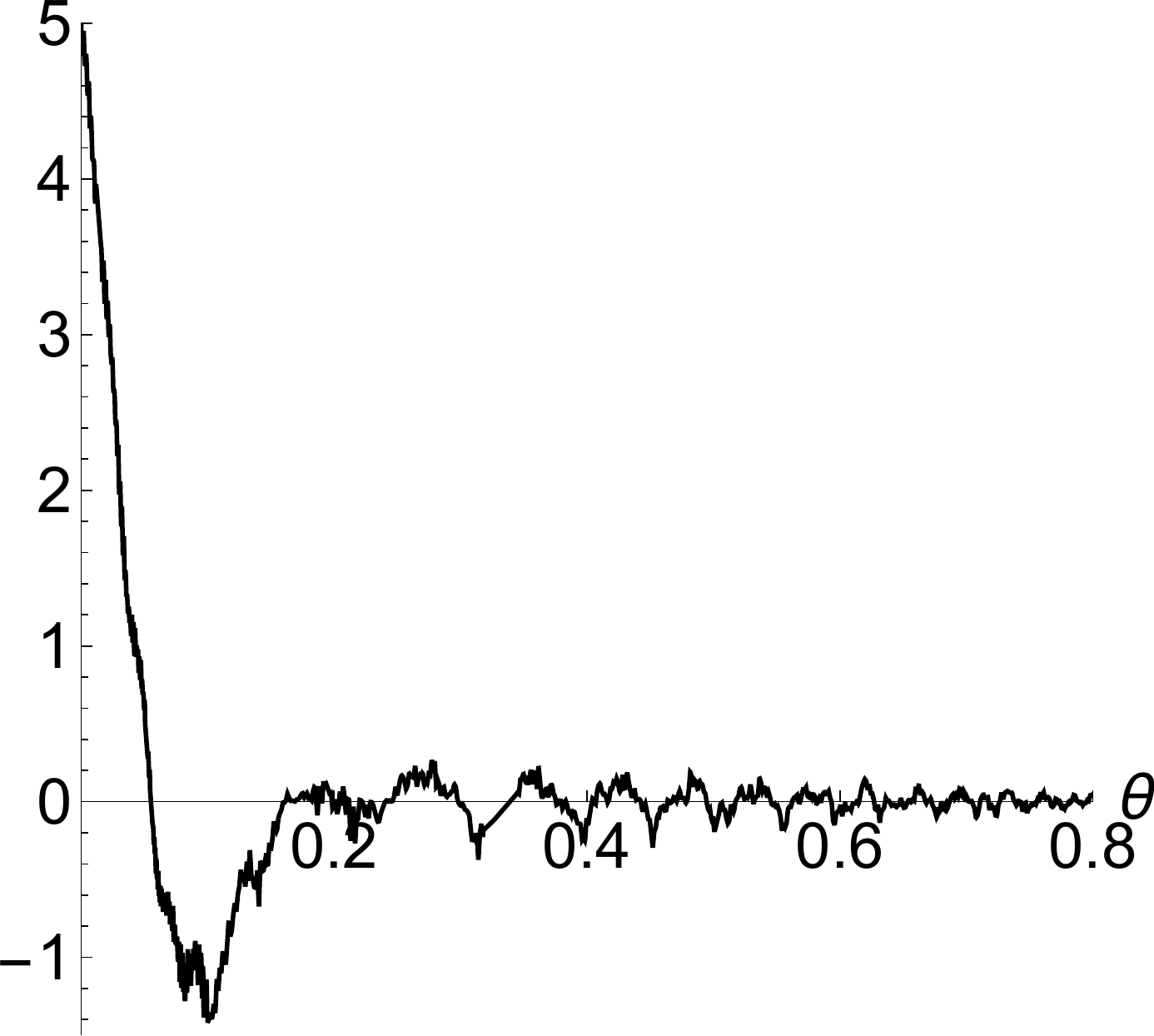}};
\node[draw] at (-2,2.75) {a};
\node at (8,0) {\includegraphics[width=6cm]{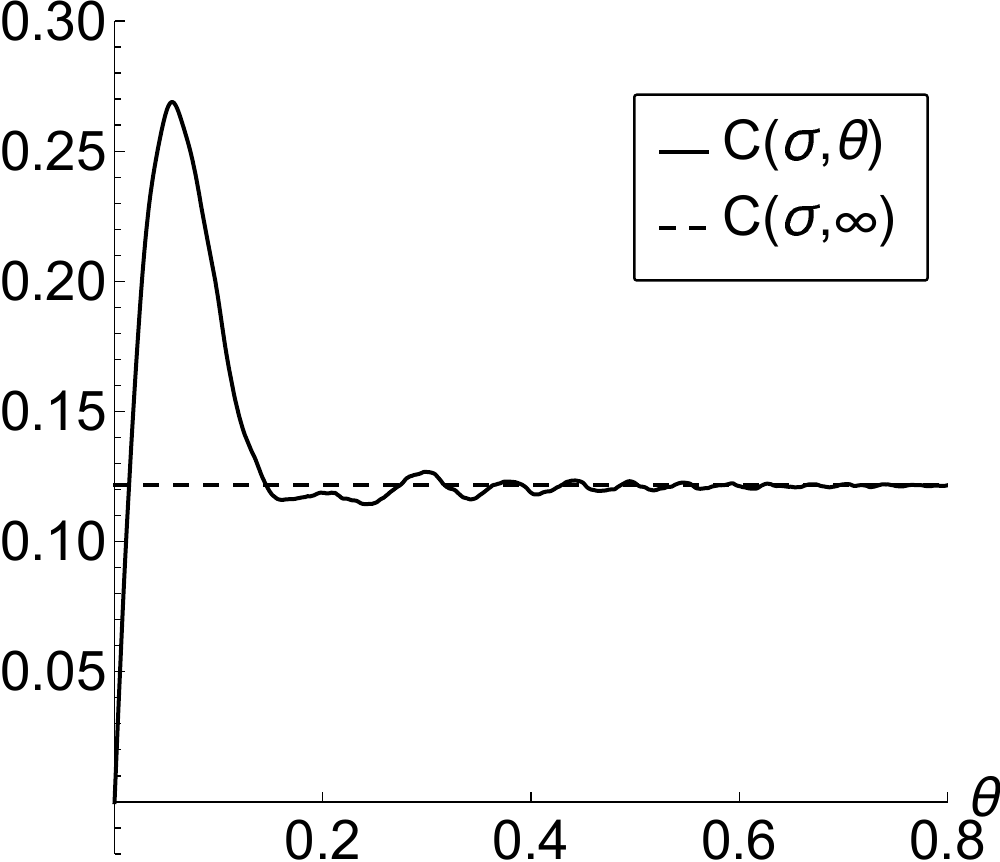}};
\node[draw] at (6.2,2.75) {b};
\end{tikzpicture}
\begin{figure}[h!!]
\centering
\caption{a) Example integrand to be integrated over to calculate $\partial \mathcal{I}_{-}/\partial \sigma$. b) The corresponding cumulative distribution function.}\label{fig:plotnum}
\end{figure}

\figref{fig:plotnum} shows the major contribution to the $\theta$-integral originates from $\theta \ll 1$ and how the frequency of oscillation increases with $\theta$.
It is instructive to compare \figref{fig:plotnum}	with a point outside the sharp edges (at $\sigma = \pm 0.5$) of the magnetic field, to illustrate differences between the full result and the LCFA. Consider the sharper magnetic field in the right-hand plot of \figref{fig:edged2} (where $\delta^{2}=0$, $\Phi=0.01$, $\xi=10$, $\eta_{k}=0.1$) for $\sigma=0.575$. According to the LCFA, there is no contribution here, but as is clear from \figref{fig:edged2}, the full probability \emph{does} actually contribute. Plotting $\partial P/\partial \sigma$ in \figref{fig:plotnum2}, we see that indeed in the full probability there is zero contribution around $\theta=0$, however, at precisely the value that $\sigma-\theta/2 < 0.5$, the integral begins to contribute with increasing $\theta$ (recall that the Kibble mass in \eqnref{eqn:Km} is defined using a window average \eqnref{eqn:wa} evaluated for phases between $\sigma \pm \theta/2$).
\newline

\begin{tikzpicture}[domain=0:4] 
\node at (0,0) {\includegraphics[width=6cm]{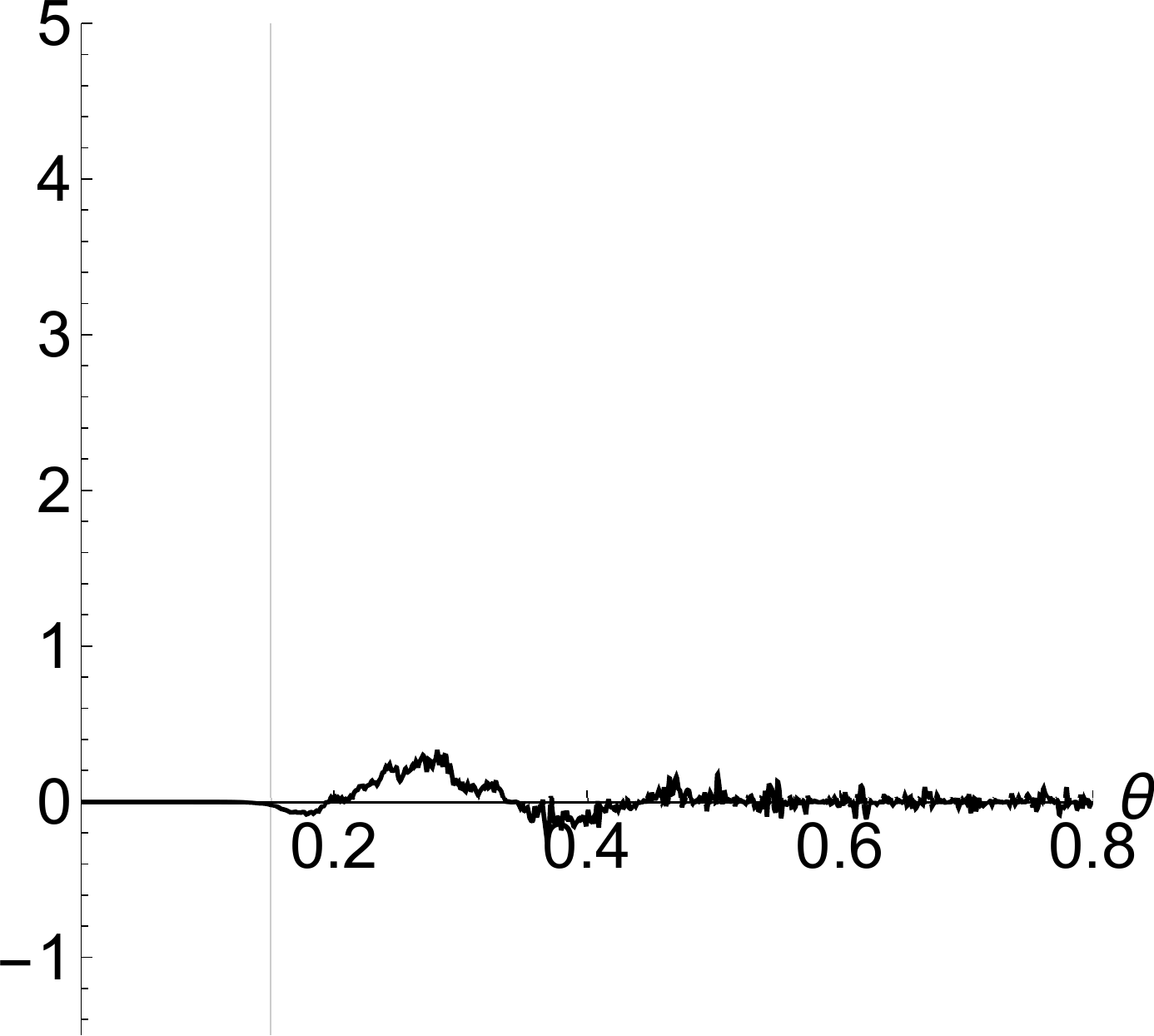}};
\node[draw] at (-2,2.75) {a};
\node at (8,0) {\includegraphics[width=6cm]{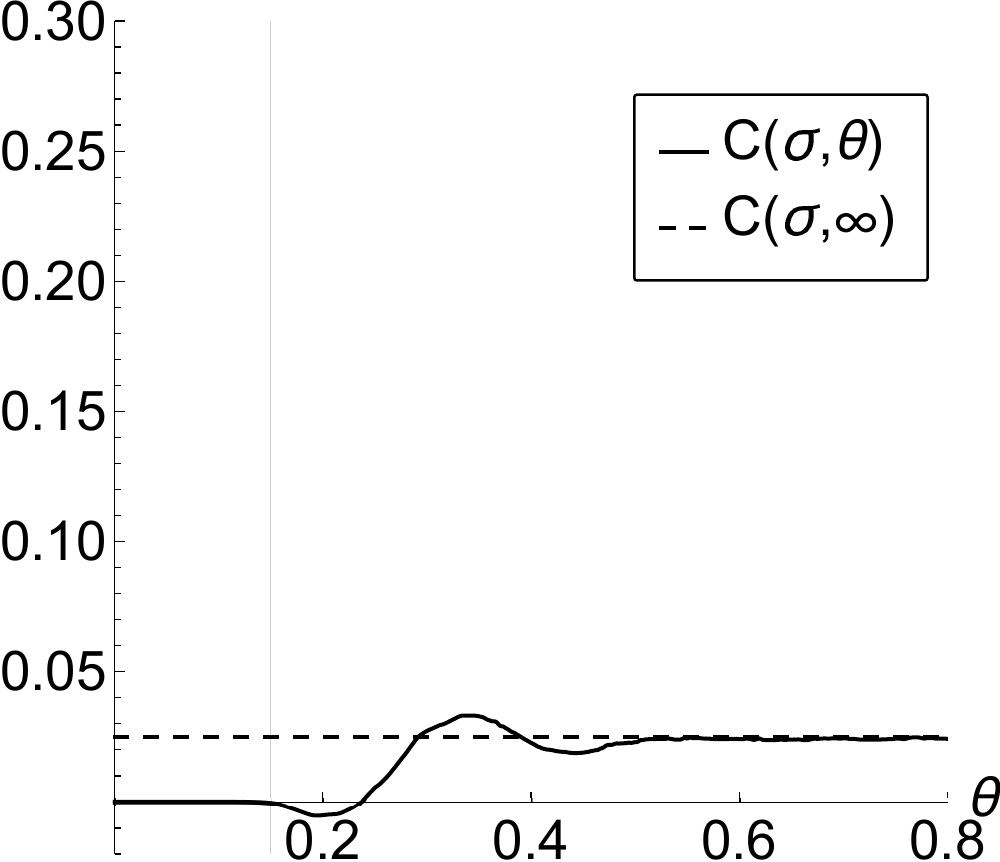}};
\node[draw] at (6.2,2.75) {b};
\end{tikzpicture}
\begin{figure}[h!!]
\centering
\caption{a) A plot of $\partial^{2}\mathcal{I}_{-}/\partial \sigma \partial \theta$ at $\sigma=0.575$ for the magnetic field with $\xi=10$, $\eta_{k}=0.1$ and $\Phi=0.01$. b) The corresponding cumulative distribution function. The vertical line in the plots is at $\theta/2=\sigma-0.5$, i.e. the distance away from the edge of the magnetic field.}\label{fig:plotnum2}
\end{figure}

The LCFA is an expansion in small $\theta$ and therefore at $\sigma=0.575$ it is not sensitive  to contributions from the field at $\sigma=0.5$, even if the field changes a substantial amount at this point.
\newline

Finally in \figref{fig:plot5}, we give some examples of the LCFA probability for pair-creation in a magnetic field in the above-threshold case. We choose $\Phi=0.2$, as this was where the numerics and LCFA agreed well for the massless case.

\begin{tikzpicture}[domain=0:4] 
\node at (0,0) {\includegraphics[width=6.4cm]{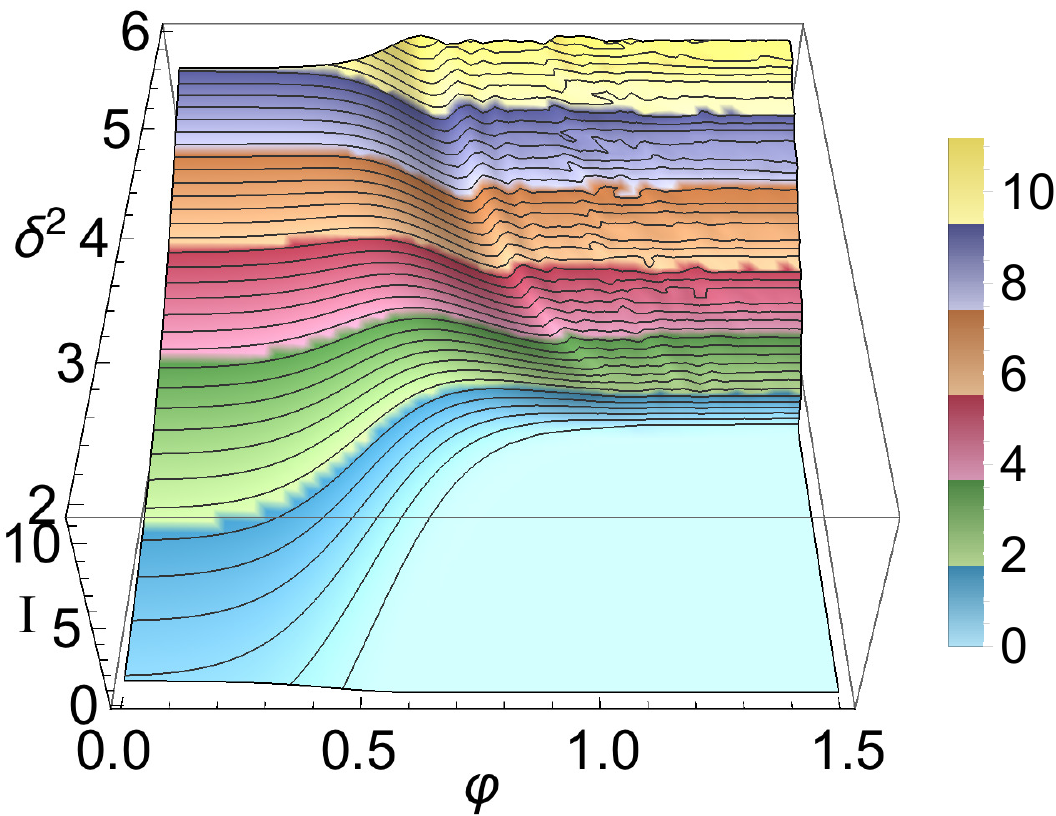}};
\node[draw] at (-3,2.5) {a};
\node at (8,0) {\includegraphics[width=7.0cm]{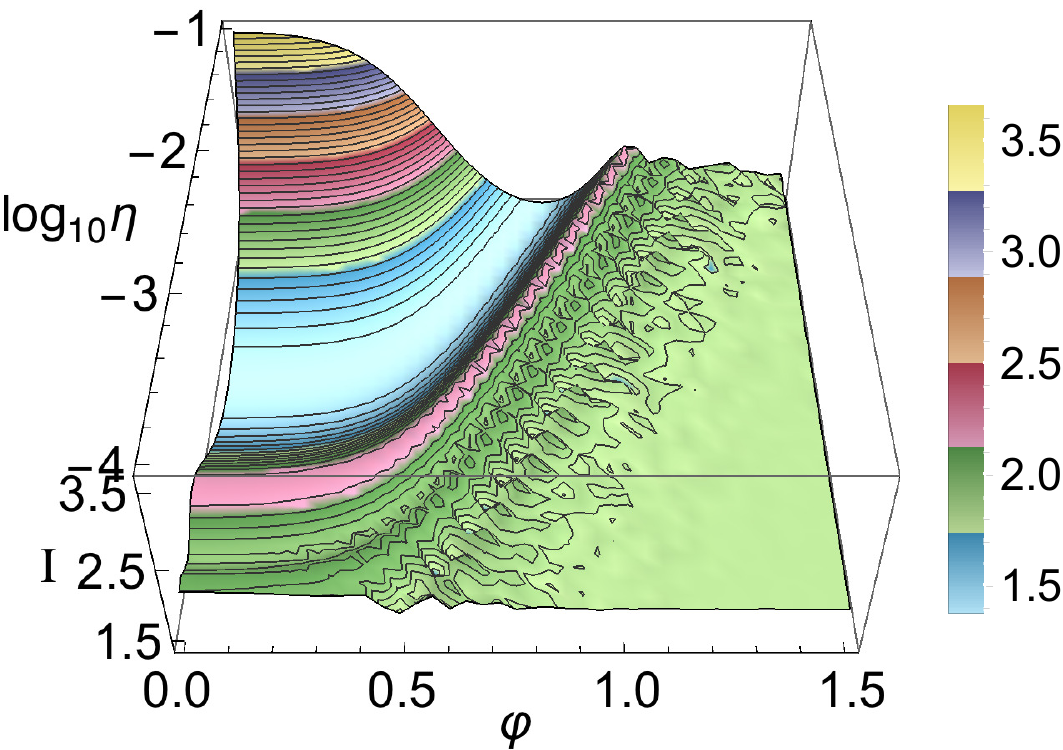}};
\node[draw] at (5,2.5) {b};
\end{tikzpicture}
\begin{figure}[h!!]
\centering
\caption{Plots for $\mathcal{I}_{\ps}^{\tsf{\tiny{LCFA}}}$ edges for the first case in \figref{fig:edged2}, with $\eta_{k}=0.1$, $\xi=10$, $\Phi=0.2$, $\delta^{2}=4.1$ (unless these parameters are varied in the plot). Left: the transition from below- to above-the-threshold behaviour. Right: the role of the energy parameter in increasing decay in the magnetic field region.
(The colours represent the vertical axis height as denoted by the scale alongside each plot.)}\label{fig:plot5}
\end{figure}

In \Ben{\figrefa{fig:plot5}} we note that, as the axion mass is increased, the effect of the magnetic field is one from increasing axion decay ($\delta^{2}\leq 4$), to the point where the magnetic field only slightly modifies the probability, and in fact in some regions of parameter space suppresses it. In \Ben{\figrefb{fig:plot5}}, we see that as the energy parameter is increased for a typical above-the-threshold scenario ($\delta^{2}=4.1$), the effect of the magnetic field enhances pair decay. As the axion impinges from the vacuum in the magnetic field, the transition in the probability from vacuum decay to field-assisted decay is nontrivial, and depending on the axion's energy, can also lead to a suppression of the decay probability.
%
%
%
%
%
%
%
\section{Summary} \label{discussion}
The decay of a massive ALP into an electron-positron pair in a high intensity EM background field has been investigated. An example derivation for the case of massive axions was presented, and the results for massive scalars and vector bosons were also given. Although the pre-exponents of scalars and vectors are different to the pseudoscalar case, the nonpertubative exponential dependency and kinematics are identical and so we expect our results to have significance for these cases as well. A constant crossed field was chosen as an example background to investigate how the decay depends on experimental and axion parameters. Three distinct regions were identified: i) below threshold decay, where the rate of decay was via tunneling through the background field; ii) above threshold decay which can proceed in the absence of a background field; iii) strong-field limit, where the concept of a threshold disappears. With interest in lab measurement in a constant magnetic field, we calculated the decay probability in a quasi-constant crossed field of finite spatial extent. This is expected to be a good approximation to decay in a constant magnetic field for highly-relativistic seed ALP particles. Using a phenomenological model, the effect of field ``edges'' and hence strong field gradients was investigated.

In below-threshold decay, a new mass-dependent tunneling exponent was identified, which shows how the gap to the threshold pair-creation energy of $2m$ is partially bridged by the mass of the axion. This is reminiscent of Schwinger pair production catalysis \cite{dunne09} by a second, higher-frequency background overlaid on a constant background, where here, the ALP mass plays the role of the higher-frequency background. The same expression for the tunnelling exponent was arrived at independently by using simple energy-momentum arguments in \cite{akhmedov11,fedotov15}.

The case of above-threshold decay shows an interplay between the two channels of: i) vacuum decay and ii) field-stimulated decay through a nonperturbative dependence on $\chi_{k}$ (a combination of external-field and ALP particle parameters). The field was found to both \emph{increase} and \emph{decrease} the probability for ALP decay, due to it inducing an oscillation in the probability around the vacuum value. This oscillation is in the ALP mass parameter, but also in $\chi_{k}$ and hence the field strength and ALP lightfront momentum. An asymptotic formula for the oscillations was found using a stationary phase analysis, and so, even if the background is not a constant crossed field, it is expected that such oscillations are a general characteristic of the above-threshold decay of ALPs into electron-positron pairs in plane-wave fields.

In the strong-field regime, $\chi_{k} \gg 1$, which would be challenging to arrive at in the lab, but may have significance in some astrophysical scenarios, the concept of a threshold disappears, and the nonperturbative asymptotic result depends only on the field. This is to be expected - eventually if the field is strong enough, the vacuum decay channel is negligibly small, and so loses meaning.

To investigate a quasi-constant crossed field of finite spatial extent in the lab, we introduced a field with dimensionless ``sharpness'' parameter, $\Phi$, parametrising the departure from a top-hat shape at $\Phi = 0$. We found that the sharper the field, the worse the approximation of taking the probability rate for a constant field and integrating it over the field shape (the so-called locally-constant field approximation). In particular, there are still contributions from outside of the field, which we found to be traceable to interference over the trajectory of the seed particle, which are absent from the simple approximation. Finally, identifying a region where the locally constant field approximation was valid, we presented the nontrivial dependency of the decay probability in a magnetic field with edge, as the mass of the decaying particle crosses the vacuum decay threshold.
%
%
%
%
%
%
%
\acknowledgments
The authors thank Subir Sarkar for suggested improvements to the manuscript. B. K. acknowledges computational resources supplied by A. Ilderton. B. K. and B. M. D. acknowledge funding from Grant No. EP/P005217/1.  B. M. D. acknowledges funding from the Slovenian Research Agency (research core funding No. P1-0035 and J1-8137).  K. A. B. and G. G. would like to thank AWE plc for support.

\bibliography{current}

\end{document}